\documentclass{PoS}

\title{The theory of Pulsar Wind Nebulae: recent progress}

\ShortTitle{Theory of Pulsar Wind Nebulae}

\author{\speaker{Elena Amato}\thanks{A footnote may follow.}\\
        INAF - Osservatorio Astrofisico di Arcetri, Largo E. Fermi, 5, 50125, Firenze, Italy;\\
        Dipartimento di Fisica e Astronomia, Universit\`a degli Studi di Firenze, Via G. Sansone 1, 50019,  Sesto F. no (Firenze), Italy.\\
        E-mail: \email{amato@arcetri.astro.it}}

\abstract{Pulsar Wind Nebulae are highly intriguing astrophysical objects in many respects. They are the brightest and closest class of relativistic sources, and hence the ultimate laboratory for the physics of relativistic plasmas: several processes observed (or inferred to occur) in other classes of relativistic sources can here be studied with unique detail, like the acceleration and collimation of relativistic outflows, or the acceleration of particles at relativistic shocks.

Here I review the current status of our theoretical understanding of Pulsar Wind Nebulae in light of the most recent 2D and 3D MHD modelling of these sources. I will discuss how these studies are taking us to the point when we can reliably use multi-wavelength observations of these nebulae as a diagnostics of the hidden physics of the pulsar wind and of the mechanism(s) through which particles are accelerated at the highly relativistic shock that terminates the wind. 

Finally I will briefly discuss recent progress in the modelling of evolved Pulsar Wind Nebulae and of the escape of particles from these systems. This effort is instrumental to credibly assess the role of Pulsar Winds as sources of cosmic ray leptons, and has recently been recognised to have important implications also on cosmic ray transport in the Galaxy.}

\FullConference{High Energy Phenomena in Relativistic Outflows VII - HEPRO VII\\
		9-12 July 2019\\
		Facultat de Fisica, Universitat de Barcelona, Spain}

\begin{document}

\section{Introduction}
\label{sec:intro}
Pulsar Wind Nebulae (PWNe hereafter) are bright nebulae of non-thermal emission powered by the wind of a fast-spinning, highly magnetised neutron star, often also detected as a pulsar. The neutron star looses most of its rotational energy in the form of a relativistic outflow, whose energy is shared between a toroidal magnetic field and a material component, mostly made of electron-positron pairs. In the immediate vicinities of the pulsar, the outflow is cold and radiatively inefficient. The bright nebular emission only starts after the flow is slowed down and a large fraction of its energy is efficiently converted into particle acceleration: this happens at a termination shock (TS hereafter) resulting from the interaction between the relativistic wind and the non-relativistic confining medium, typically the ejecta of the parent supernova explosion for young objects ($\lesssim$ few $\times 10^4$ yr) and the ram pressure exerted by the Interstellar Medium (ISM hereafter) for older ones (see e.g. \cite{gaenslerslane06} for a review of the different structures and evolutionary stages). 

The class prototype, the Crab Nebula, is one of the best studied objects in the Universe and its emission has been measured over more than 20 decades in frequency, from radio wavelengths to very high energy gamma-rays (see e.g. \cite{meyer10} for a recent collection of all available data and \cite{hawccrab} for the most recent high energy measurements). The main radiation mechanism is synchrotron emission with a spectrum extending up to photon energies $\gtrsim 100$ MeV, while at higher energies Inverse Compton scattering becomes important. Detection of both synchrotron and Inverse Compton emission allows one to disentangle the properties of magnetic fields and particles: interpretation of the Crab Nebula spectrum requires the presence of a rather intense magnetic field (0.1-1 mG) and of a particle distribution in the form of a broken power-law, namely $E^{-\alpha}$, with $\alpha\approx 1.5$ for $E\lesssim$ 300 GeV and $\alpha\approx 2.3$ for 300 GeV < E < 1PeV (\cite{volpi08}). The Crab Nebula is the only source in which we have direct evidence of particle acceleration up to PeV energy, namely the energy at which the cosmic ray spectrum steepens and the highest energy that we think achievable in Galactic sources. Moreover the total energy radiated by the nebula corresponds to an extraordinarily large fraction of the pulsar spin-down luminosity (>25\%). These two facts make the Crab Nebula the most efficient accelerator known in the Galaxy. At the same time, the mechanism by which such efficient particle acceleration is achieved is most mysterious. 

In fact, as we will discuss in the following, the particle spectrum that we observe in the Crab Nebula cannot result (or at least not primarily) from diffusive shock acceleration (e.g. \cite{blandeich}), the most commonly invoked process for particle acceleration in Astrophysics. A number of alternative acceleration mechanisms have been proposed, but these also are not free of difficulties, and/or their viability depends on properties of the pulsar wind that are not straightforward to constrain. In the following I will review recent efforts aimed at constraining the three most important unknowns that are relevant for particle acceleration: the pulsar multiplicity, namely the number of electron-positron pairs created in the magnetosphere by each electron extracted from the star surface; the pulsar wind magnetisation, namely the ratio between magnetic and kinetic energy in the pulsar wind; the possible presence of ions in the pulsar wind.

These properties of pulsar winds have broad implications that go well beyond the physics of PWNe. The pair multiplicity is deeply connected with the workings of the pulsar magnetosphere, and might help unveiling some of them \cite{arons12}. On the other hand, the presence of hadrons in pulsar winds would make these sources potential important contributors of hadronic cosmic rays at the high energy end of the galactic spectrum (E>100 TeV) and lend support to scenarios that consider young, highly magnetised pulsars as primary sources of Ultra High Energy Cosmic Rays (e.g. \cite{kotera15} and references therein).

In recent years, pulsars and their nebulae have already entered the realm of sources that are interesting for cosmic ray physics, thanks to the detection of the so-called positron excess \cite{pamela,ams02}. Indeed, the unexpected rise with energy of the positron/electron ratio in the flux of cosmic rays detected at the Earth, at energies larger than $\sim$ 30 GeV, finds in the contribution of PWNe a most natural explanation \cite{bykov17}. This contribution would primarily come from evolved nearby systems and its assessment, in terms of intensity and spectrum, is fundamental to put constraints on the properties of cosmic ray transport in the vicinities of the solar system and also to determine whether it leaves room for further components of more exotic nature, either astrophysical sources with peculiar properties or dark matter related phenomenology \cite{crrev18}.  

Most of the particle release by PWNe is expected to occur at late stages of their evolution, after the pulsar has left its parent supernova remnant and is moving through the ISM. Establishing the spectrum of pairs that these sources release in the Galaxy requires knowledge of their dynamics and evolution until late stages and of the processes that connect the nebular plasma with the outer medium and govern the particle escape. 

In the following, I will review how our understanding of PWN physics has improved during the last 15 years thanks to detailed modelling of the dynamics and emission of these sources, mainly carried out by means of multi-D relativistic MHD simulations and computation of emission maps on top of those. I will describe recent efforts towards constraining the pulsar wind parameters and the mechanisms of particle acceleration. Finally I will discuss the latest work devoted to the description of evolved systems and of the properties of particle release from them.

\section{The pulsar wind}
\label{sec:wind}
In the last two decades, much insight has been gained on the structure of the pulsar wind thanks to a combination of analytical and numerical modelling. The wind emanating from the pulsar magnetosphere is thought to be highly magnetised in the vicinity of the pulsar light cylinder, and well described as a force-free outflow. At larger distances, the flow streamlines become asymptotically radial and the embedded magnetic field largely toroidal. The wind energy flux is maximum in the equatorial plane of the pulsar rotation and then progressively decreases with latitude towards the pole, scaling $\propto \sin^2\theta$ \cite{bogov99,spit06} (or more rapidly \cite{philippov18}) with polar angle $\theta$. The equatorial region of the wind is characterised by the presence of an oscillating current sheet extending within an angular sector of width equal to twice the inclination between the magnetic and rotation axis of the pulsar. When this part of the outflow impacts the TS, oppositely directed magnetic field lines are forced to come close together and driven magnetic reconnection is likely to lead to complete dissipation of the alternating component of the field \cite{kirkrev}. Hence, at the TS, we expect the magnetic field strength to peak at intermediate latitudes, and decrease both towards the pole and the equator. Recent numerical simulations of the wind have shown that also the particle number density has a complex dependence on latitude \cite{philippov18}. 

The wind properties that are important to determine what particle acceleration process can be at work are: its magnetisation, $\sigma$, the pair multiplicity, $\kappa$, and the wind baryon load. In general these will very with latitude along the shock front, while their integral is constrained by the total wind luminosity, which equals, to a very good approximation, the rotational energy lost by the pulsar per unit time:
\begin{equation}
\dot E_{\rm R}=\kappa \dot N_{\rm GJ} m_e c^2 \Gamma_{\rm wind} \left(1+\frac{\xi m_i}{\kappa m_e}\right)(1+\sigma)\ ,
\label{eq:winden}
\end{equation}
where $c$ is the speed of light, $m_e$ is the electron (and positron) mass, $m_i$ is the mass of the ions possibly present in the wind, $\xi$ is a parameter, with $0<\xi<1$, defining the extraction rate of ions in terms of the {\it Goldreich \& Julian rate} \cite{gj69}, $\dot N_{\rm GJ}$. The latter is the rate at which electrons are extracted from the pulsar by the surface electric field: it only depends on the pulsar period $P$ and magnetic moment $\mu$, and can be written as $\dot N_{\rm GJ}=3 \times 10^{30} \mu_{30} P^{-2}$, where $P$ is in seconds and $\mu_{30}$ in units of $10^{30}$ G cm$^{3}$. Finally, the last parameter entering Eq.~\ref{eq:winden} is the wind magnetisation, $\sigma$, defined as the ratio between the wind Poynting flux and particle kinetic energy:  
\begin{equation}
\sigma=B^2/(4 \pi n_e m_e \Gamma^2 c^2)\ ,
\label{eq:sigma}
\end{equation}
with $B$ the magnetic field and $n_e$ the comoving electron and positron number density. 
As we will discuss in the following, from modelling of the Crab Nebula synchrotron emission, we infer the following values of the parameter at the TS: $10^4<\Gamma_{\rm wind} <10^7$, $10^3<\kappa<10^6$, $\sigma\lesssim 1$-few. Concerning this last parameter, it is worth noticing that its best estimate has changed drastically since the first attempts at MHD modelling of PWNe \cite{kc1}: the 1D steady state MHD description was in fact leading to estimate $\sigma=3\times 10^{-3}$, about 3 orders of magnitude lower than what current 3D MHD models require, and much work was devoted to understand the processes by which such a low value could be reached at the TS starting from $\sigma\approx 10^4$ at the light cylinder. 

In the following we discuss how these values of the various quantities come about and impact the viability of the proposed mechanisms.

\section{Particle acceleration mechanisms in PWNe}
\label{sec:accel}
We already mentioned that a broken power-law spectrum, changing from harder to softer than $E^{-2}$ at around 300 GeV, is thought to be typical of PWNe \cite{nicjon}. While distributed acceleration may take place in these sources, as we will later discuss, and is sometimes invoked in connection with GeV particles (e.g. \cite{lyutikov19}), acceleration of particles to TeV energies and above is generally believed to occur at the wind TS. As we discuss below, this is the most relativistic shock in Nature, with a Lorentz factor estimated in the range $10^3<\Gamma_w<10^7$. Particle acceleration is very difficult to explain at all in this context, even with maximum energies and efficiencies much lower than the extraordinary values we infer for the Crab Nebula.

Among the various mechanisms that have been proposed, the three best studied ones, and the ones we will mostly discuss, are: (i) diffusive shock acceleration, or 1$^{\rm st}$ order Fermi mechanism; (ii) driven magnetic reconnection; (iii) resonant absorption of ion-cyclotron waves. 

\subsection{Diffusive acceleration at a relativistic shock}
\label{sec:fermiI}
Diffusive shock acceleration is the most commonly invoked particle acceleration mechanism in Astrophysics. The basic idea behind the process resides in the fact that particles gain energy every time they cross a shock front, both from upstream to downstream and vice-versa. This mechanism has been shown to guarantee efficient particle acceleration in the context of non-relativistic shock waves \cite{caprioli14}, and also at relativistic unmagnetised shock waves \cite{spit08}, but its performance at relativistic magnetised shocks is very poor \cite{sironi15}. In fact, the essential requirement for Fermi process to work is that particles are able to cross the shock many times. This condition is very difficult to satisfy at a relativistic magnetised shock: in fact, unless the shock normal and the magnetic field direction are aligned within an angle $1/\Gamma$, the shock is effectively superluminal and particles have a very hard time returning from downstream. In order for them to successfully fight advection and diffuse back to the shock, a very high level of turbulence is required, with $\delta B/B\gg 1$. This requirement is easy to satisfy only if the magnetisation is very low, in which case the growth of Weibel instability ensures a sufficiently high level of turbulence and efficient particle acceleration. In terms of the wind magnetisation, the condition for this mechanism to work is $\sigma\lesssim 10^{-3}$. In this case the shock can in principle accelerate particles, with a nearly universal spectrum $\approx E^{-2.3}$, similar to what we infer from X-ray observations of PWNe. However, it is worth noticing that even when this condition is satisfied, the turbulence that develops is typically small-scale turbulence. This implies that the acceleration time increases with $E^2$, where $E$ is the particle energy. Therefore the particles cannot be accelerated to very high energies. This is a general problem with invoking Fermi acceleration at relativistic shocks (e.g. \cite{sironi15}). A possible way to overcome this issue would be offered by the existence of large scale turbulence of external origin in the shock vicinity, for instance MHD turbulence induced by the corrugation of the TS could be an interesting candidate in the case of PWNe \cite{lemoine16}.

\subsection{Driven magnetic reconnection}
\label{sec:drivrec}
We mentioned above that the equatorial sector of the wind with alternating field lines is likely to undergo magnetic reconnection when compressed at the shock. Aside from creating a low magnetisation region where Fermi mechanism could in principle operate, reconnection can be itself accompanied by very efficient particle acceleration: if conditions are appropriate, the entire magnetic energy that is dissipated can be converted into particle acceleration, leading to very hard ($E^{-\alpha}$ with $1<\alpha<2$) and extended power-law spectra. The flat spectral indices typical of the process are in the range that one generally infers from radio emission of PWNe. Moreover this mechanism is able to completely erase the Maxwellian component of the particle distribution, which on the contrary is always present in the case of Fermi acceleration and never observed in PWNe. 

However, the outcome of the process in terms of spectral slope and extension depends on the initial magnetisation and pair loading of the flow. Detailed PIC simulations have shown that in the case of a particle spectrum with $\alpha\approx 1.5$, such as observed in the Crab Nebula, in order to account for the 3 decades in particle energy that radio emission spans, one needs $\sigma\gtrsim 30$ and $\lambda/(r_L \sigma)\gtrsim$ a few tens, where $\lambda=2 \pi R_{\rm LC}$ is the wavelength of the stripes ($R_{\rm LC}$ is the light cylinder radius) and $r_L$ is the particle Larmor radius at the TS \cite{sironi11}. Assuming energy conservation along the streamlines, the latter condition can be rephrased as $\kappa\gtrsim\sigma/(1+\sigma)(R_{\rm TS}/R_{\rm LC})$, where $R_{\rm TS}$ is the TS radius. In the case of the Crab Nebula, considering the TS position coincident with the boundary of the underluminous region surrounding the pulsar at optical wavelengths, $R_{\rm TS}\approx 0.1 pc$, which implies $\kappa>10^8$, much larger than current pulsar theories can account for \cite{timokhin19}, and also much larger than inferred from observations, as we will later discuss. 

\subsection{Resonant cyclotron absorption in a ion-doped plasma}
\label{sec:rca}
An alternative mechanism that has been proposed to explain particle acceleration at relativistic transverse shock is that based on resonant absorption by electrons and positrons of the cyclotron radiation emitted by ions that are also part of the flow (e.g. \cite{hoshino91}). This mechanism works for whatever $\sigma$, but requires that most of the energy of the pulsar wind be carried by ions. The basic physical picture is as follows: at the crossing of the TS, the sudden enhancement of the magnetic field sets the previously drifting plasma into gyration. The leptons quickly thermalise through emission and absorption of cyclotron waves, but ions with the same initial Lorentz factor (the wind is cold, so that all particles were moving with the same bulk Lorentz factor) react on time-scales that are longer by a factor $m_i/m_e$. If the wind is sufficiently cold ($\delta u/u<m_e/m_i$, with $u$ the four velocity) before the TS, the ions emit waves with large power not only at the fundamental frequency of their gyration, but up to a frequency $m_i/m_e$ times higher, which can then be resonantly absorbed by the pairs. The resulting acceleration efficiency $\epsilon_{\rm acc}$, spectral slope $\alpha$ and maximum energy $E_{\rm max}$, all depend on the fraction of energy carried by the ions $U_i/U_{\rm tot}$. PIC simulations show a wide variety of values: $\epsilon_{\rm acc}= few(30)\%$, $\alpha > 3(< 2)$, $E_{\rm max}/(m_i \Gamma c^2) = 0.2(0.8)$ for $U_i/U_{\rm tot} = 0.6(0.8)$ \cite{amatoarons}. Once again the pulsar multiplicity - and, as we will see, the related question of the origin of radio particles - plays a crucial role.

\section{Lessons from multi-D MHD simulations}
\label{sec:mhd}
In the last 2 decades, our understanding of PWN physics and the possibility of constraining pulsar wind parameters has received great benefit from numerical studies in relativistic MHD. The era of multi-D MHD simulations was triggered in this field by {\it Chandra} X-ray images, showing axisymmetric structures, in the form of jets and torii, in virtually all spatially resolved PWNe \cite{gaenslerslane06}. The X-ray appearance was especially impressive in the case of the Crab Nebula, where the polar jet was clearly seen to originate from much closer to the pulsar than the inferred position of the TS in the equatorial plane. The difficulties at explaining collimation within the ultra-relativistic pulsar wind were a primary driver of the idea that the outflow could be anisotropic, with an energy flux larger at the equator than at the pole, so as to produce a TS much closer to the central star along the polar axis than in the equatorial plane \cite{lyub02}. Axisymmetric relativistic MHD simulations \cite{komlyub04,ldz04} proved indeed that the above mentioned split monopole solution for the pulsar wind structure naturally results in a non-spherical TS. The numerical experiments also showed that for proper values of the wind parameters (anisotropy and magnetisation) emission structures similar to those observed in X-rays arise \cite{ldz06}. In particular the existence of a jet allowed to put a lower limit on the wind average magnetisation: only for high enough values of $\sigma$, the magnetic field downstream reaches equipartition and its tension is able to divert the flow towards the polar axis giving rise to a jet. The required value of $\sigma$ depends on the nebular expansion velocity and in the case of the Crab Nebula the requirement is found to be $\sigma\gtrsim 0.03$ \cite{ldz06}, an order of magnitude larger than estimates based on 1D MHD modelling \cite{kc1}. 

In spite of the success at explaining the X-ray morphology of the Crab Nebula in fine details, including jets, rings and the X-ray knot, 2D MHD modelling soon proved inadequate to fully account for the actual plasma dynamics in PWNe \cite{volpi08}. Its shortcomings became immediately apparent when computing the nebular Spectral Energy Distribution. In particular, the value of $\sigma$ that best reproduces the nebular morphology fails to reproduce the multi-wavelength spectrum of the Crab Nebula: the average value of the nebular magnetic field is too low, and as a consequence, if one considers injection of a sufficiently large number of particles, so as to reproduce the high energy synchrotron spectrum, the Inverse Compton emission is largely overestimated. On the other hand, larger values of $\sigma$ lead to exceedingly large hoop stresses: too large a fraction of the flow is diverted towards the pole and the equatorial emission is too faint \cite{olmi16}. 

\begin{figure}
\begin{center}
\includegraphics[width=0.47\textwidth]{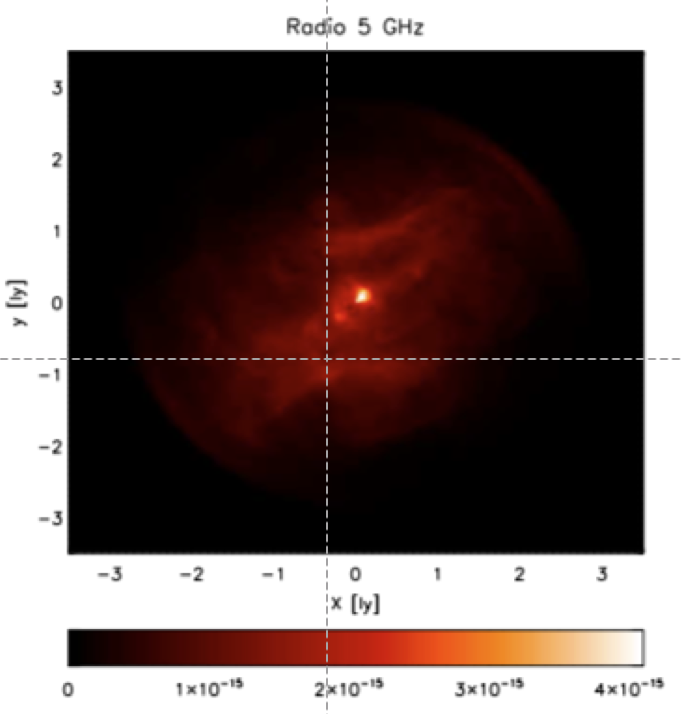}
\includegraphics[width=0.46\textwidth]{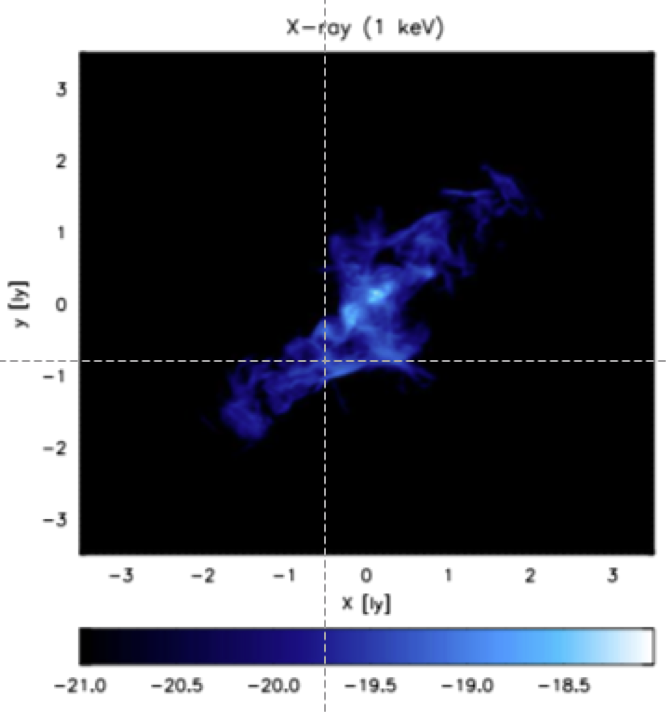}
\caption{Emission maps of the Crab Nebula at 5 GHz (left panel) and 1keV (right panel) resulting from a 3D simulation with $\sigma=1$ extending up to 500 yr after the SN explosion (see \cite{olmi16} for detailes). Surface brightness is in units of mJy arcsec$^{-2}$ and the spatial scale is in ly. }
\label{fig:crab3d}
\end{center}
\end{figure}

The main ingredient that the axisymmetric description is forcedly missing is the development of kink type instabilities, whose possible relevance had already been suggested in the context of explaining how to confine a wind with an initially high magnetisation at the shock \cite{begelman98}. Indeed, when the first 3D MHD simulations of these systems were finally performed \cite{porth13}, it was immediately clear that the same values of the parameters produce very different nebulae in 2D and 3D. In this latter, more realistic treatment, effective magnetic reconnection takes place downstream of the shock, reducing the hoop stresses and allowing to reproduce the nebular morphology with a larger initial value of $\sigma$. Of course in terms of reproducing both the spectrum and the morphology what is needed is a reorganisation of the field, rather than simple dissipation, so as to decrease the field tension while still keeping its strength. Long duration simulations \cite{olmi16,porth17}, in which an asymptotic self-similar solution appears to be fully reached, show that after an initial phase of continuous decrease, the average field strength reaches a constant value. However, up to now no simulation is available that can fully reproduce both the morphology and SED of the Crab Nebula. Fig.~\ref{fig:crab3d} shows the radio (5 GHz, left panel) and X-ray (1 keV, right panel) emission from Crab resulting from a 3D simulation with $\sigma=1$ extended up to 500 yr after the explosion. One can notice that the main issues are with the X-ray morphology: the brightness contrast between the X-ray torus and the jet is clearly too low, and the equatorial region is too compact.

One firm conclusion that can be drawn from these studies is that the wind magnetisation of the shock must be much larger than previously thought: in the case of the Crab Nebula one may estimate $\sigma\gtrsim$ a few. This has profound implications on the viable acceleration processes at the shock: as already discussed, the Fermi process cannot operate in regions where the plasma has $\sigma>0.001$. Of course at the TS the local magnetisation of the flow will depend on latitude above the equator and substantially lower values can be found in the equatorial sector that hosts the current sheet. However when we evaluate  the fraction of the flow that satisfies the condition $\sigma<0.001$, we find it to be insufficient to account for the Crab Nebula X-ray emission \cite{amato14}. This conclusion might be changed if effective focusing of particle trajectories towards the equator occurs and the level of magnetic turbulence downstream of the shock is high enough \cite{giacinti18}.

\begin{figure}
\begin{center}
\includegraphics[width=0.47\textwidth]{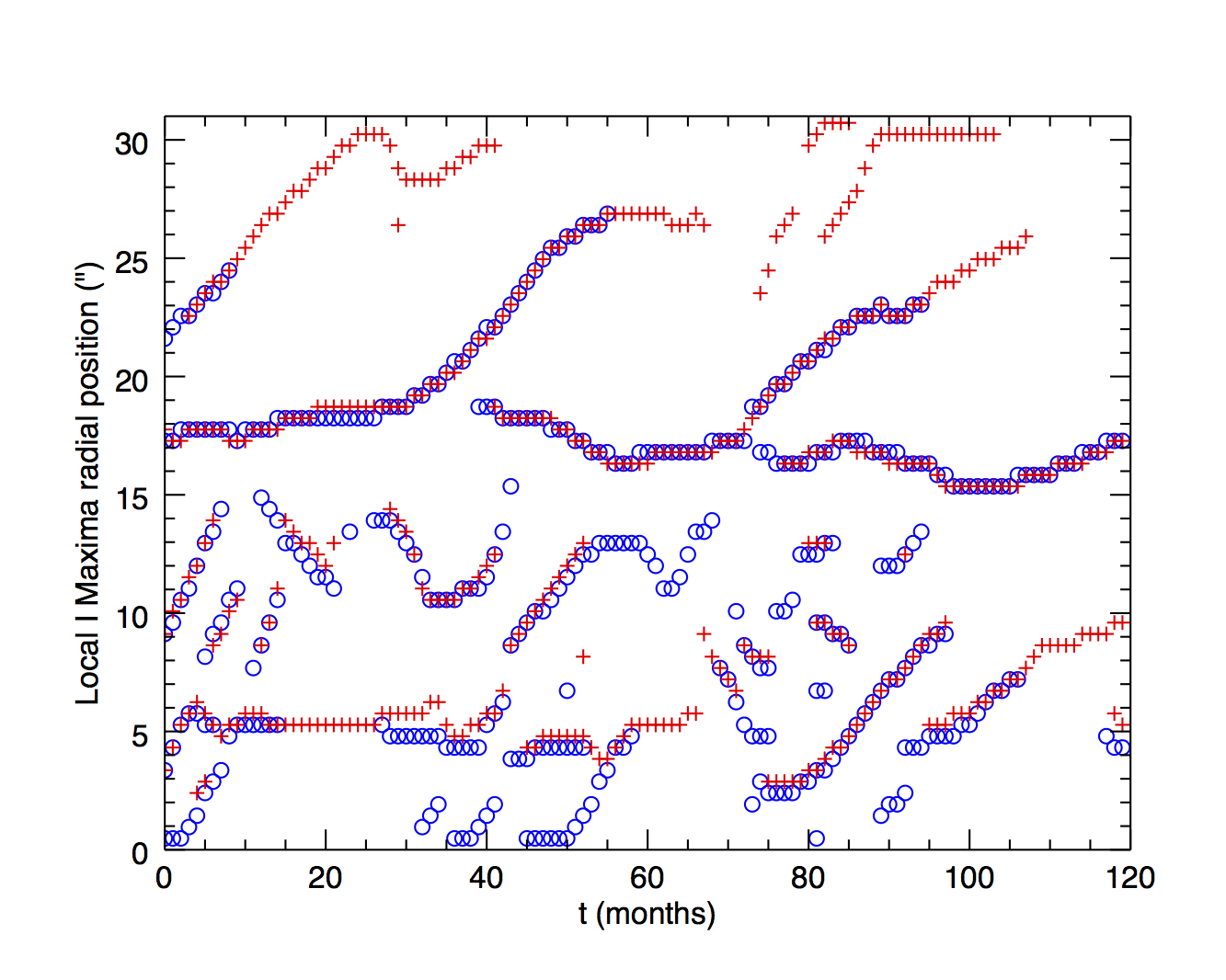}
\includegraphics[width=0.47\textwidth]{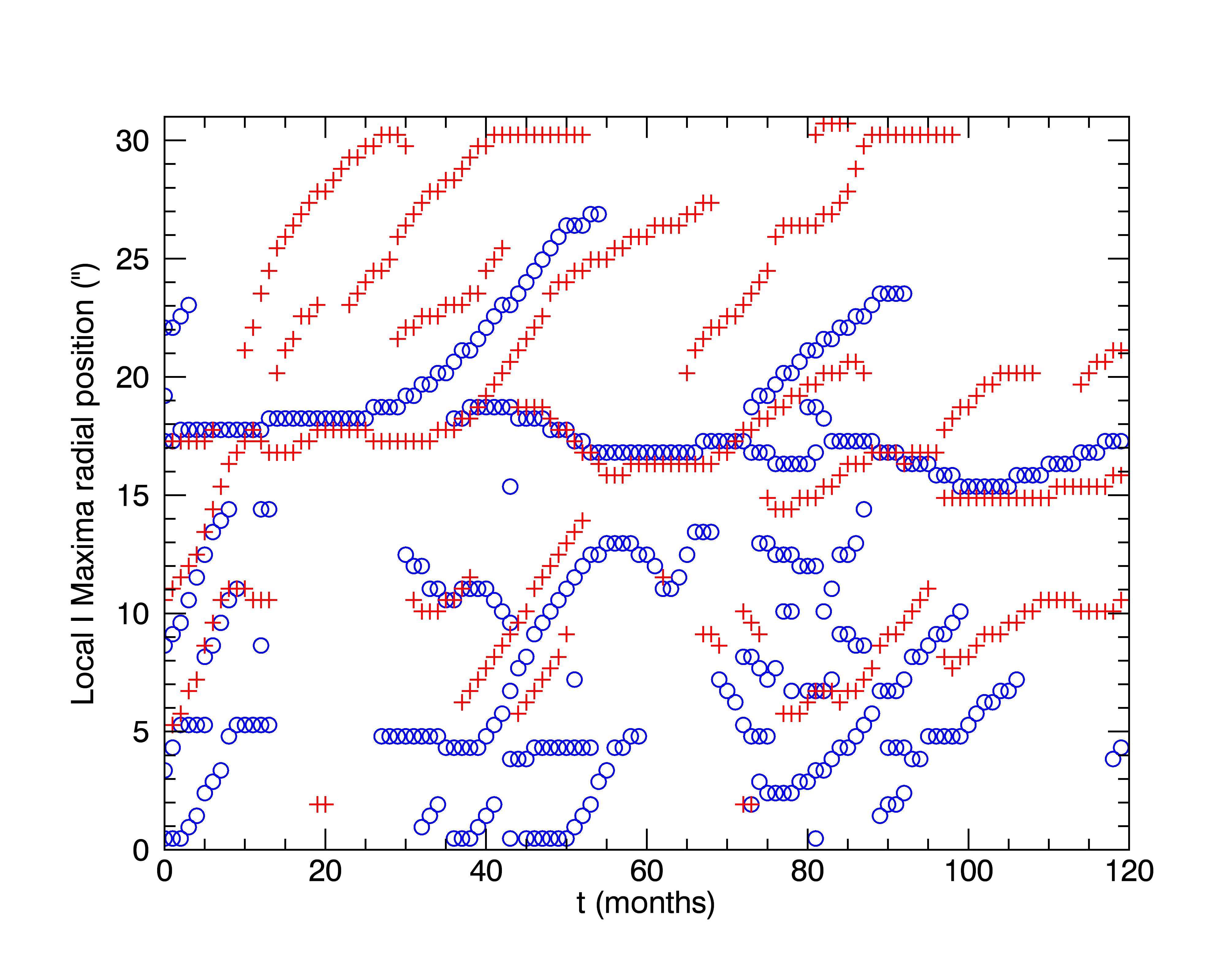}
\caption{Time variability of the {\it wisps} at radio (red) and X-ray (blue) frequencies. The plot on the left is for a scenario in which both the radio and X-ray emitting particles are injected uniformly around the shock front. In the plot on the right radio emitting particles are still injected everywhere along the shock front, but X-ray emitting ones are only injected in the region $0^\circ<\theta<30^\circ$.}
\label{fig:wisps}
\end{center}
\end{figure}

In any case, the possibility that different acceleration mechanisms operate at different latitudes along the TS is an interesting one to consider. This kind of investigation is primarily prompted by observations, aside from any theoretical consideration. In fact, \cite{schweizer13,bieten04} performed an investigation of the so-called {\it wisps}, the time variable features in the inner regions of the Crab Nebula, at multi-wavelength. The result of such analysis is that the wisps are not coincident at radio, optical and X-ray frequencies. Within a MHD description of the flow, the {\it wisps} are interpreted as regions of enhanced magnetic field and fast flow motion towards the observer (increasing the Doppler boosting effect). Therefore, differences in their appearance and behaviour can only be accounted for if the particles responsible for the emission at different frequencies are injected at different locations along the shock front\footnote{Notice that this is still true even if the {\it wisps} are interpreted as magnetic compressions induced by ions \cite{spitarons}}. In particular, detailed analysis of synthetic emission maps, computed on top of the MHD flow with different prescriptions on the injection of particles in different energy ranges, showed that the observed {\it wisps} behaviour is best reproduced if optical/X-ray emitting particles are injected mostly at low latitudes and radio emitting particles are injected either at higher latitudes or everywhere along the shock front \cite{olmi15}. This fact is illustrated in Fig.~\ref{fig:wisps}: in the plot on the left, where injection is uniform across the shock front at all energies,  the {\it wisps} are mostly coincident at radio and X-ray frequencies, except for the effect of synchrotron losses, that make X-ray {\it wisps} disappear more quickly with distance from the pulsar; in the plot on the right, on the other hand, where X-ray emitting particles are only injected in the vicinity of the equator and radio emitting ones everywhere along the TS, one can find situations in which the structures in different wave bands are misaligned, in space and time. 

In fact the observed behaviour of the {\it wisps} at radio wavelengths is even compatible with a more extreme hypothesis, namely that the particles responsible for radio emission are uniformly distributed in the nebula rather than accelerated at the shock front and then advected with the flow. Available radio images of the Crab Nebula do not allow to discriminate between a scenario in which radio emitting particles are accelerated at the TS alone and one in which they are uniformly distributed throughout the nebula \cite{olmi14}. The only scenario that radio data allow us to exclude is one in which the particles responsible for the emission are fully relic, namely injected in the nebula at early times (as suggested e.g. by \cite{atoyan99}) and then advected with the flow.  A relic origin of the low energy particle population is instead compatible with observations if one assumes that they are not simply advected, but rather effectively scattered and reaccelerated by turbulence (Fermi II type acceleration) or magnetic reconnection. 

This latter result has profound implications on the estimate of the pulsar multiplicity, $\kappa$. Indeed, for typical particle spectra in PWNe, radio emitting particles are dominant by number. If they are part of the pulsar wind, they determine the value of $\dot N$ that must be supplied by the wind, and hence the values of $\Gamma$ and $\kappa$ in Eq.~\ref{eq:winden}. In the case of the Crab Nebula one finds $\dot N=10^{40}\ {\rm s}^{-1}$, and hence $\kappa=10^6$ and $\Gamma=10^4$. On the other hand, if these particles are not currently part of the pulsar wind, but were rather injected in the nebula at some earlier time (when the pulsar was much younger and more energetic \cite{atoyan99}) or even extracted from the thermal plasma in the supernova ejecta, then the flow parameters would be determined by the population of X-ray emitting particles. These require $\dot N=10^{38.5} \ {\rm s}^{-1}$, and hence $\kappa=10^4$ and $\Gamma=$few$\times 10^6$.

These estimates offer different constraints on the viable mechanisms of particle acceleration at the TS. If radio emitting particles are part of the pulsar wind, the estimated value of $\kappa$ is larger than current models of pair creation in the magnetosphere can account for \cite{timokhin19}, yet below the value ($10^8$) that would allow acceleration by driven magnetic reconnection to produce a particle spectrum such as that observed in Crab \cite{sironi11}. However, it should be kept in mind that $\kappa>10^8$ was derived based on particle density at the TS position in the equatorial plane (see \S~\ref{sec:drivrec}). We now know that the TS is much closer to the star along the polar axis, and since the density scales $\propto r^{-2}$ (with $r$ the distance from the pulsar), a value $\kappa\lesssim 10^6$ would be sufficient if reconnection was taking place there. The problem is that we expect high latitude stripes only in the case of an orthogonal rotator.

If $\kappa\approx10^6$, the pairs are so many that, even if present in the wind at the level of a Goldreich \& Julian flux, ions could not dominate its energy content and resonant cyclotron absorption of ion waves could not work as an acceleration mechanism for pairs \footnote{Notice, however, that this conclusion assumes the same Lorentz factor for pairs and ions. A recent suggestion by \cite{coroniti17} is that the two components might be separated in the wind and the ions have a much larger Lorentz factor than the pairs, dominating the flow for any plausible value of $\kappa$}.
On the other hand, if radio emitting particles are not being supplied by the pulsar, $\kappa$ is within the range of values that magnetospheric models predict and is also such that ions can be energetically dominant and accelerate pairs at the shock. It is interesting to notice that the presence of ions in the pulsar wind would also ease the conversion of Poynting flux to kinetic energy flux that must occur in the wind between the light cylinder and the TS \cite{kirk19}.

These would be PeV ions, particles at the end of the galactic CR spectrum. Their presence might show up in $\gtrsim$100 TeV gamma-rays and neutrinos, possibly detectable by upcoming facilities for multi-messenger astronomy \cite{amato03}. These highly energetic ions, when Eq.~\ref{eq:winden} is rescaled for the parameters appropriate to describe newly born (and hence fast spinning) highly magnetised ($B_{\rm psr}>10^{13}$ G) neutron stars, would make them suitable candidates for the yet unknown origin of Ultra High Energy Cosmic Rays (e.g. \cite{kotera15}).

\section{Particle escape from evolved systems}
\label{sec:bow}
Aside from the connection just highlighted, PWNe have recently called the attention of the CR physics community as possible primary contributors of cosmic ray positrons. In the last decade, it has become evident, with ever increasing statistical significance, that the ratio between cosmic ray positrons and electrons increases with energy above $\sim$10 GeV \cite{pamela, ams02}. This finding is in contrast with the expectations of models considering positrons as pure secondary products of cosmic-ray interactions during propagation through the Galaxy and has prompted many efforts to provide an explanation \cite{crrev18}. A pulsar related origin of the unexpected positrons is among the most popular suggestions. In particular, evolved PWNe seem a most promising candidate \cite{bykov17}. 

The {\it positron excess} is detected at the energies that characterise the radio emitting particles in these systems and the excess is best explained with a spectrum $E^{-1.2/-1.5}$ which is what typically inferred from radio emission of PWNe. These particles, however, have small Larmor radii compared to the size of the system and cannot easily escape as long as the PWN is young and inside its parent SNR. However the high average proper motion of the pulsar population causes the nebula to leave the parent SNR within a time-scale of order few $\times 10^4$ yr, and at that point the nebula is likely to become a Bow Shock PWN. At that stage, particles are not only free to leave the PWN from the tail of the bow shock, but also likely to escape from close to the bow shock head, thanks to the development of instabilities that break the contact discontinuity and let the relativistic plasma flow freely in the ISM. Proof of such a phenomenon is directly observed at least in a few systems showing extended trails of X-ray emission developing perpendicular to the bow shock \cite{huibecker,pavan14,pavan16}.

In fact, recent observations have cast some doubts on the effective escape of leptons from the pulsar surroundings. Indeed the detection by HAWC \cite{hawcpos} of extended haloes of multi-TeV emission around Geminga and PSR B0656+14 leads to infer that the diffusion coefficient in the vicinity of these systems is much reduced with respect to the galactic average, and particles undergo relevant energy losses due to ICS before leaving the region and becoming part of the galactic cosmic ray pool. In light of all this, it has become clear that assessing the amount and the spectrum of leptons released by PWNe is mandatory in order to make progress on cosmic ray origin and galactic transport. 

\begin{figure}
\begin{center}
\includegraphics[width=0.3\textwidth]{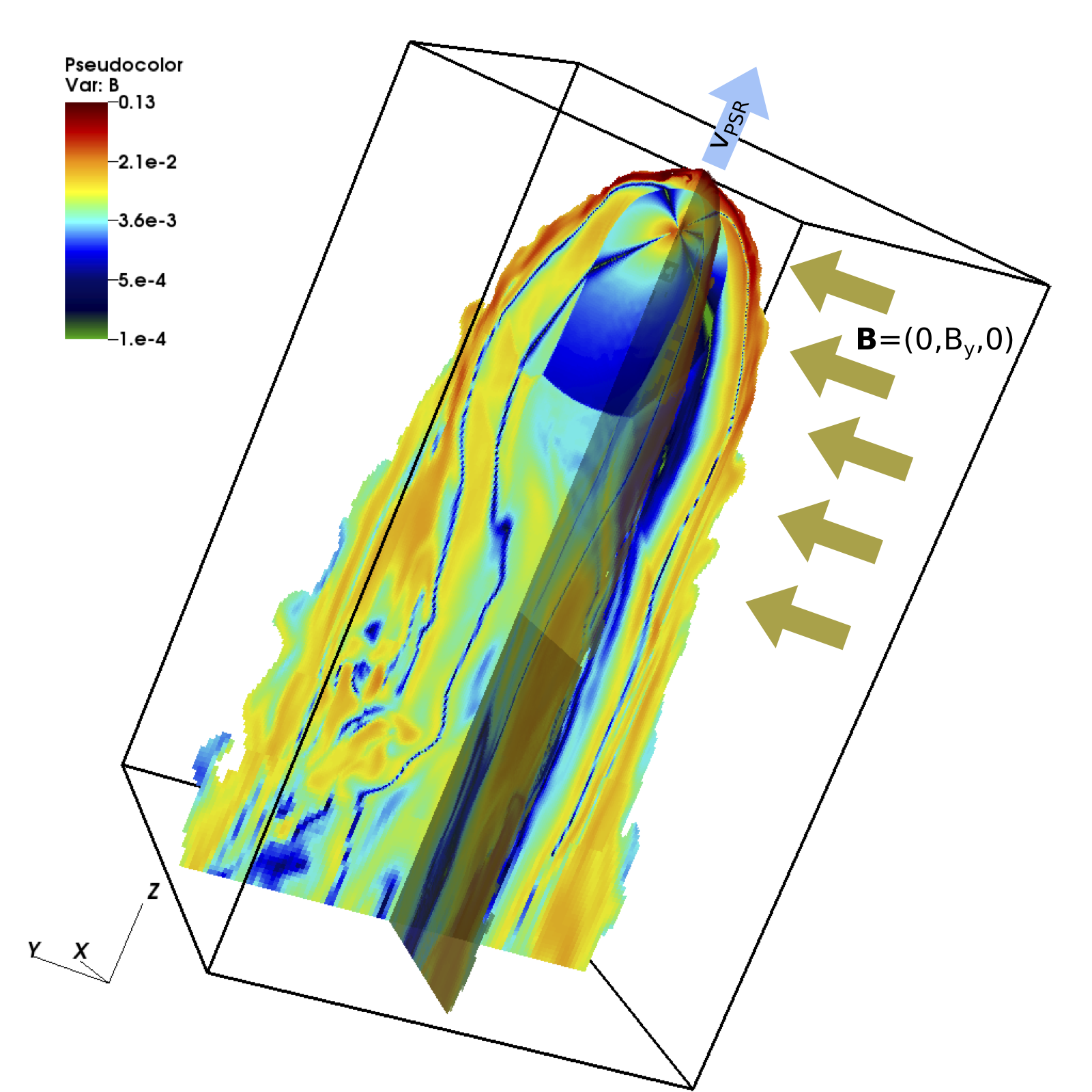}
\includegraphics[width=0.3\textwidth]{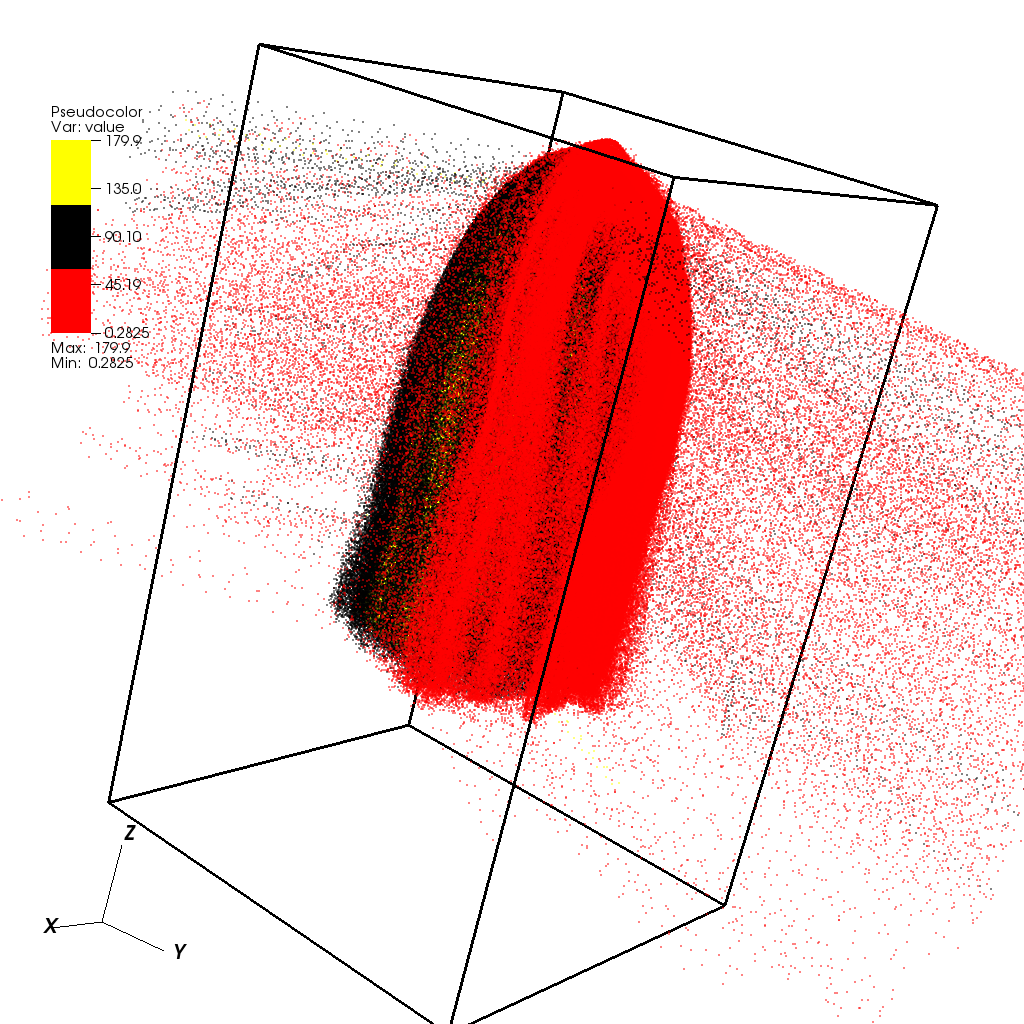}
\includegraphics[width=0.3\textwidth]{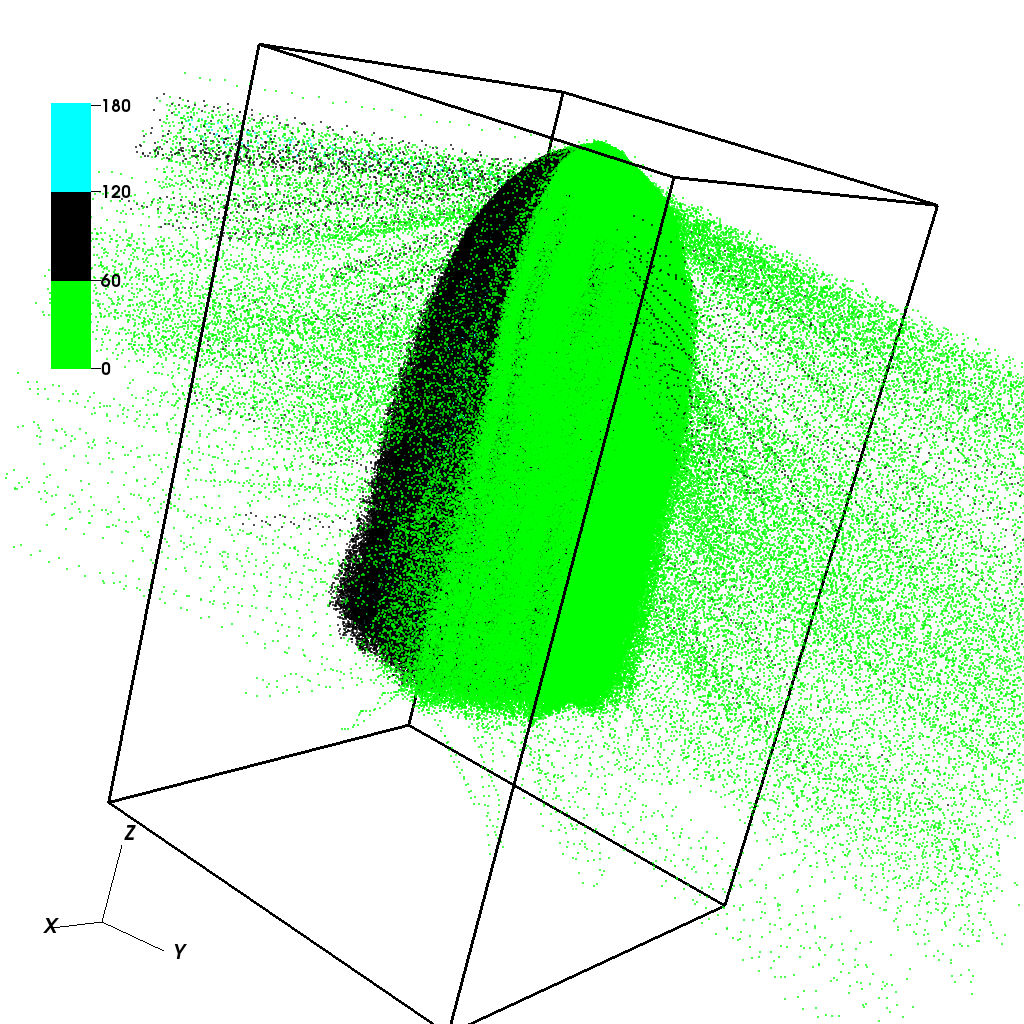}
\caption{3D simulations of a Bow Shock PWN \cite{olmi20}. The pulsar wind magnetisation is $\sigma=1$ and the pulsar spin axis is inclined by $45^\circ$ with respect to the pulsar proper motion. The 3 panels show: magnetic field strength (left); escaping positrons (centre); escaping electrons (right). The colour scale for the latter panels refers to the latitude along the shock front at which the escaping particles were initially injected.} 
\label{fig:bow}
\end{center}
\end{figure}

Some recent work on the subject has been carried out in \cite{olmi19a,olmi19b}, where an extended set of 3D relativistic MHD simulations of Bow Shock PWNe was been performed with the aim of better understanding the dynamics of these systems and the mechanisms of particle release in the ISM. A very important result of this study is that particles at energies close to the pulsar potential drop ($E\approx e \sqrt{\dot E/c}$) can escape from these nebulae not only from the back, but also from the sides of the bow shock (see central and right panel in Fig.~\ref{fig:bow}), just as observed in the Guitar and Lighthouse nebulae \cite{huibecker,pavan14,pavan16}. The escape occurs along the magnetopause at the contact discontinuity between the PWN and the shocked ISM (left panel of Fig.~\ref{fig:bow}). Depending on the inclination of the pulsar spin axis with respect to the direction of its proper motion, positrons and electrons escape preferentially from different sides of the nebula \cite{olmi19b}. This opens the possibility for the growth of non-resonant modes of the streaming instability \cite{bell04,amatoblasi09}, which could lead to effective confinement of the pairs in the surrounding of the nebula \cite{olmi20}. This could provide an explanation for the extended gamma-ray haloes observed by HAWC \cite{hawcpos}, which are otherwise difficult to explain as self-generated by the particles themselves \cite{evolimorlino}.

\section{Summary \& Conclusions}
In this article I have tried to review the current status of our theoretical understanding of PWN physics, mostly focusing on the aspects that are most relevant to clarify the question of how the most relativistic shocks hosted by these sources can accelerate particles so efficiently. 3D MHD simulations indicate that $\sigma\gtrsim$ few is required to reproduce the spectrum and morphology of the Crab Nebula. As a result the 1$^{\rm st}$ order Fermi process can only operate in a tiny sector of the TS around the equator, which doubtfully encloses a large enough part of the outflow. However the equatorial region is where X-ray emitting particles must come from, and maybe subtle kinetic effects and large scale MHD turbulence are playing an important role to enhance the efficiency of the process. Of the other proposals, driven magnetic reconnection is really difficult to make viable, while the possibility that ions are part of the flow and responsible for the acceleration of pairs appears still completely open. The presence of relativistic ions would even contribute to clarify some controversial aspects of the evolution of the wind before the TS. The upcoming gamma-ray and neutrino facilities might finally prove or disprove it, also contributing to clarify the role of pulsars as sources of Ultra High Energy Cosmic Rays.

In the final part of this review I described recent work on Bow Shock PWNe, a subset of PWNe that is attracting growing interest in the cosmic ray physics community, as possible major contributors of cosmic ray leptons. The modelling of these sources is at a very early stage, but preliminary studies of the particle escape have highlighted very interesting aspects: particles are likely released as an electrically charged current, that might induce powerful instabilities and drastically change the properties of transport around the source. Further work is needed to clarify how these recent results compare with gamma-ray observations of TeV haloes and how they affect the overall contribution of PWNe to the galactic pool of cosmic ray leptons.

\acknowledgments{I thank R. Bandiera, N. Bucciantini and B. Olmi for continuous discussions on these subjects and for providing the images in Fig.~\ref{fig:bow}. This work was partially funded through Grants ASI/INAF n. 2017-14-H.O and SKA-CTA-INAF 2016}

\end{document}